\documentclass[aps,twocolumn,prb,floatfix,showpacs,superscriptaddress]{revtex4-1}
\usepackage[columnwise]{lineno}
\usepackage{graphicx}
\usepackage{float}
\usepackage{subfigure}
\usepackage{dcolumn}
\usepackage{gensymb}
\usepackage{bm}
\usepackage{epstopdf}
\usepackage{subfigure}
\usepackage{amsmath}
\usepackage{color}
\usepackage{units}
\usepackage{amssymb}
\newcommand{\etal}{\textit{et al.\/}}

\usepackage{color}
\usepackage[usenames,dvipsnames,svgnames,table]{xcolor}
\usepackage{hyperref}

\hypersetup{
     colorlinks   = true,
     citecolor    = blue,
     linkcolor    = blue
           }
\newcommand{\amend}{\textcolor{black}}
\newcommand{\amendyk}{\textcolor{black}}

\begin{document}

\title{Magnetism and ultra-fast magnetization dynamics of Co and Co-Mn alloys at finite temperature}



\author{R.~Chimata}
\email{raghuveer.chimata@physics.uu.se}
\affiliation{Department of Physics and Astronomy, Material Theory, University Uppsala, SE-75120 Uppsala, Sweden}

\author{E.~K.~Delczeg-Czirjak}
\affiliation{Department of Physics and Astronomy, Material Theory, University Uppsala, SE-75120 Uppsala, Sweden}

\author{A.~Szilva}
\affiliation{Department of Physics and Astronomy, Material Theory, University Uppsala, SE-75120 Uppsala, Sweden}

\author{R.~Almeida}
\affiliation{Department of Physics and Astronomy, Material Theory, University Uppsala, SE-75120 Uppsala, Sweden}
\affiliation{Faculdade de F\'{i}sica, Universidade Federal do Par\'{a}, Bel\'{e}m, PA, Brazil}

\author{Y.O.~Kvashnin}
\affiliation{Department of Physics and Astronomy, Material Theory, University Uppsala, SE-75120 Uppsala, Sweden}

\author{M.~Pereiro}
\affiliation{Department of Physics and Astronomy, Material Theory, University Uppsala, SE-75120 Uppsala, Sweden}

\author{S.~Mankovsky}
\affiliation{Department of Chemistry, University of Munich, Butenandtstrasse 5-13, D-81377 Munich, Germany}

\author{H.~Ebert}
\affiliation{Department of Chemistry, University of Munich, Butenandtstrasse 5-13, D-81377 Munich, Germany}

\author{D.~Thonig}
\affiliation{Department of Physics and Astronomy, Material Theory, University Uppsala, SE-75120 Uppsala, Sweden}

\author{B.~Sanyal}
\affiliation{Department of Physics and Astronomy, Material Theory, University Uppsala, SE-75120 Uppsala, Sweden}

\author{A.~B.~Klautau}
\affiliation{Faculdade de F\'{i}sica, Universidade Federal do Par\'{a}, Bel\'{e}m, PA, Brazil}

\author{O.~Eriksson}
\affiliation{Department of Physics and Astronomy, Material Theory, University Uppsala, SE-75120 Uppsala, Sweden}

\date{\today}

\begin{abstract}
\noindent
Temperature-dependent magnetic experiments like pump-probe measurements generated by a pulsed laser have become a crucial technique for switching the magnetization in the picosecond time scale. Apart from having practical implications on the magnetic storage  technology, the research field of ultrafast magnetization poses also fundamental physical questions. \amend{To correctly describe the time evolution of the atomic magnetic moments under the influence of a temperature-dependent laser pulse, it remains crucial to know if the magnetic material under investigation has magnetic excitation spectrum that is more or less dependent on the magnetic configuration, e.g. as reflected by the temperature dependence of the exchange interactions. In this article, we demonstrate from first-principles theory that the magnetic excitation spectra in Co with fcc, bcc and hcp structures are nearly identical in a wide range of non-collinear magnetic configurations.
This is a curious result of a balance between the size of the magnetic moments and the strength of the Heisenberg exchange interactions, that in themselves vary with configuration, but put together in an effective spin Hamiltonian results in a configuration independent effective model. We have used such a Hamiltonian, together with ab-initio calculated damping parameters, to investigate the magnon dispersion relationship as well as the ultrafast magnetisation dynamics of Co and Co-rich CoMn alloys.}
\end{abstract}


\maketitle

\section{Introduction}
Ultrafast magnetism, with relevant time-scales being an order of a few pico seconds, has become an intense research field. The motivation may be found both in fundamental aspects as well as practical implications of these phenomena. Most of the information stored technologically is done in a magnetic medium. Hence, the possibility to write and retrieve information in a magnetic material at a high speed and with low energy consumption has obvious societal implications. For this reason, ultrafast magnetization dynamics has naturally become an intense research field. The experiment by Beaurepaire and co-workers\cite{beaurepaire} represents a breakthrough experiment, with several experimental studies that followed.\cite{kimel1,kimel2,bigot,tudosa,bigot2,stamm,vahapalar} However, despite several years of intense investigations, a microscopic understanding of the processes of ultra-fast magnetization dynamics is far from been established.

The most common experimental technique is by pump-probe, where an optical laser pulse excites the electron sub-system. The excited electrons become thermalized quickly\cite{rhie-1}, and the thermal energy of the electron sub-system is transferred to the spin- and lattice sub-systems. This defines three thermal reservoirs, and typically the three reservoirs reach thermal equilibrium after some 10-20 pico seconds. The time evolution of the temperatures of these reservoirs may be quantified by the so called three temperature model.\cite{beaurepaire,koopmans1,koopmans2} It should be noted that in the first few pico seconds of a pump-probe experiment, the material is not in thermal equilibrium between the three reservoirs, but after sufficiently long time after the pump pulse, the temperature is the same in the different sub-systems.

On the theoretical side it has been argued that atomistic spin-dynamics simulations should be relevant over a time-scale of pico seconds and longer.\cite{book} The argument here is that a description of atomistic moments is relevant, and that these moments evolve in time with a spin-temperature that is given by the three-temperature model.\cite{raghuveer} In this description, the magnetic moments and all parameters of an effective spin-Hamiltonian are evaluated from first principles theory. Coupled to the equations of motion of the atomic spins,\cite{antropov} this allows for numerical results of the time evolution of the magnetic moments, forming an ab-initio theory that does not rely on experimental results as input. The dominating parameters of such a spin-Hamiltonian are the size of the atomic moments coupled to the inter-atomic exchange interaction.\cite{sasha} To mention an example of the fruitfulness of this approach we note that the first experimental result of fcc Ni, was reproduced by such simulations with good accuracy.\cite{evans}

Recently, it was shown that the inter-atomic exchange interactions of bcc Fe has a distinct temperature dependence, and only good agreement with experimental room temperature magnon excitations was achieved when the exchange parameters were evaluated at room temperature.\cite{attila} This demonstrates that if a too broad temperature interval needs to be covered, bcc Fe is not an ideal Heisenberg system, and that the normal concepts of a Heisenberg Hamiltonian, e.g. magnons, can still be considered although configuration dependent exchange parameters must be evaluated and used. This puts high demands if a three-temperature model is attempted to be used to reproduce an experimental pump-probe experiment, since at each time-step the exchange parameters (and magnetic moment) should in principle be recalculated, in order to take the changing temperature of the spin-system into account. If a material is a good Heisenberg system or not, i.e. if the exchange parameters are independent on the temperature or not, is difficult to stipulate before a first principles calculation of the configurational dependent exchange parameters have been made, but several systems have by now been suggested to have exchange parameters that depend more or less strongly on temperature. Hence, it seems that there are indeed very few materials that are good Heisenberg systems. In this article we demonstrate that Co, in fcc, bcc and hcp form, is rather unique in this sense, at least among the \amend{elemental metals, displaying the features of a Heisenberg magnet in a wide range of magnetic configurations.
As we shall see below, this comes with a twist, since both the values of the magnetic moments and the strength of the Heisenberg exchange depend on configuration, but put together in a spin Hamiltonian they form a model that curiously is configuration independent. We also investigate the magnetization dynamics of this system, and compare it to a Co-Mn alloy in the bcc and bct structures.} 

The article is organized as follows. In Sec.~\ref{methods} we present the theoretical tools used to describe the ground state and dynamical properties. Sec. \ref{numerical} contains the numerical details of the calculations. Results and discussions are presented in Sec. \ref{result}. Finally, we give conclusions in Sec. \ref{conclusion}.
\section{METHODS}\label{methods}
In order to investigate ultrafast demagnetization dynamics of Co and Co-Mn alloys, we combined first principles electronic structure calculations with atomistic spin dynamics simulations. These methods are described below.

\subsection{Electronic structure calculations}

The ground state electronic structure and magnetic properties of the studied materials are obtained via density functional theory (DFT) calculations. The Kohn-Sham equations are solved within the Korringa-Kohn-Rostoker (KKR) Green's function formalism as implemented in the spin-polarized relativistic KKR (SPR-KKR) package\cite{sprkkr}, and within linear muffin-tin orbital method\cite{lmto-orig} in atomic sphere approximation (LMTO-ASA). 
\amend{We used both real-space (RS)\cite{anders} and reciprocal-space\cite{py-lmto} realisations of LMTO-ASA}. 
The relativistic effects are considered by solving the fully relativistic Dirac equation \cite{dirac}. The substitutional disorder is treated by making use of the coherent potential approximation (CPA) \cite{cpa}. The high temperature paramagnetic phase was modelled by disordered local moment (DLM) approximation \cite{Pindor,Staunton} in combination with  CPA. Here the Co-Mn binary systems were treated as quaternary (Co$_{0.5}^\uparrow$Co$_{0.5}^\downarrow$)$_{1-x}$(Mn$_{0.5}^\uparrow$Mn$_{0.5}^\downarrow$)$_{x}$ alloys, with a random mixture of the two magnetic orientations of Co and Mn. DLM approach is believed to accurately describe the high temperature paramagnetic phase \cite{Pindor}, therefore we apply this tool, in case of alloys, to evaluate whether the local magnetic moments and magnetic coupling constants are sensitive to the temperature induced fluctuations or not.

The interatomic exchange interactions, \textit{J}$_{ij}$, are calculated via the Liechtenstein-Katsnelson-Antropov-Gubanov formalism {(LKAG) \cite{sasha} as implemented in the SPR-KKR and the RS-LMTO-ASA codes} and its extension to non-collinear spin arrangement \cite{attila}, (see Section \ref{sec_attila}). The site and element resolved Gilbert damping parameters ($\alpha$, $\alpha_{\rm Co}$, $\alpha_{\rm Mn}$) are calculated based on the linear response formalism  \cite{EMKK11}, (see Section \ref{sec_sergey}).

\subsection{Calculation of the interatomic $J_{ij}$ exchange parameters}\label{sec_attila}

Interatomic magnetic exchange interaction parameters, $J_{ij}$, are calculated from first principles. For collinear atomic spin alignment, the method of infinitesimal spin rotation was derived almost thirty years ago\cite{sasha}. The energy (grand potential) variation can be calculated when the atomic spin is rotated by a small angle simultaneously, and mapped onto a {bilinear spin model}:
\begin{equation}
\mathcal{H} =-{1 \over 2}\sum_{i\neq j}J_{ij}\vec{e}%
_{i}\cdot\vec{e}_{j}  \;,
\label{spinHam}
\end{equation}
where the unit vector $\vec{e}_{i}$ ($\vec{e}_{j}$) denotes the direction of the spin at site $i$ ($j$). 
Although it might seem trivial, we write also this spin-Hamiltonian in the more common form, that explicitly describes the coupling of atomic spin moments, $\vec{m}$: 
\begin{equation}
\mathcal{H} =-{1 \over 2}\sum_{i\neq j}{\tilde J}_{ij}\vec{m}%
_{i}\cdot\vec{m}_{j}  \;.
\label{spinHam2}
\end{equation}
For the discussion of our results (below) it becomes relevant to make a distinction between Eq.~\eqref{spinHam} and \eqref{spinHam2}, and the fact that $J_{ij}$ and ${\tilde J}_{ij}$ differ only by a factor $\vec{m}_i=m_i\vec{e}_i$.

The {LKAG interatomic exchange} formula can be written as $J_{ij}=A_{ij}^{\uparrow \downarrow}$, where the symbol $\uparrow$ and $\downarrow$ refer to the up and down spin channels, respectively, while
\begin{equation}
A_{ij}^{\alpha \beta }=\frac{1}{\pi }\int\limits_{-\infty }^{E_{F}}d\varepsilon \operatorname{Im}  \mbox{Tr}_{L}\left(p_{i}T_{ij}^{\alpha
} p_{j}T_{ji}^{\beta }\right) \;.  \label{AImdeff}
\end{equation}
{For collinear spin configuration the corresponding $T_{ij}^{\uparrow}$ and $T_{ij}^{\downarrow}$ matrices denote the component of the scattering path operator (SPO), $\tau_{ij}$, in the two spin channels between site $i$ and $j$ while $p_{i}$ and $p_{j}$ stand for the (spin-part) of the inverse of the} one-site scattering matrix\cite{sasha}. \amend{ In order to treat alloys or alloy analogy models, the defect atom $\mu$ is created at site $i$ by a defect matrix $D_i^\mu$ . This defect matrix is considered in the effective CPA medium $\tilde{\tau}_{i\mu,j\nu}=D_i^\mu\tau_{ij}^{\text{CPA}}D_j^\nu$ \cite{Bottcher}. Hence, the scattering path operator $\tilde{\tau}_{i\mu,j\nu}$ replaces related components of the scattering path operator in Eq.~(\ref{AImdeff}). }

In non-collinear spin arrangement, the SPO matrix elements can be grouped into a charge and spin part with the help of the two times two unit matrix and the Pauli matrices. Hence, one can define the exchange matrix $A_{ij}^{\alpha \beta }$ where indices $\alpha $ and $\beta $ run over $0$, $x$, $y$ or $z$. By using trace properties, the general symmetry relation $A^{\alpha\beta}_{ij}=A^{\beta \alpha}_{ji}$ was found. In the absence of spin-orbit coupling we can write that $\left(T^{\alpha}_{ij} \right)^{T}=T^{\alpha}_{ji}$, which implies that $A^{\alpha\beta}_{ij}=A^{\beta \alpha}_{ij} $, i.e., the A-matrix is symmetric. The grand potential (pairwise) variation is proportional to the variation of the integrated density of states, which is determined by using the Lloyd formula \cite{lloyd}. This leads to the expression
\begin{equation}
\delta E_{ij}=-2 J^{nc}_{ij} \delta \vec{e}_{i}\delta \vec{e}
_{j}-4\sum_{\mu ,\nu =x,y,z}\delta e_{i}^{\mu }A_{ij}^{\mu \nu }\delta
e_{j}^{\nu } \; \label{gen}
\end{equation}%
when the spin-orbit interaction is not considered, where
\begin{equation}
J_{ij}^{nc}=A_{ij}^{00}-A_{ij}^{xx}-A_{ij}^{yy}-A_{ij}^{zz}  \;.
\label{Jncol}
\end{equation}%
At low temperature, where the degree of non-collinearity between atomic spins is small (a regime we denote the quasi collinear regime) the second term of Eq. (\ref{gen}) does not give a significant contribution, hence we will here resolve $J^{nc}_{ij}$ for different systems, which can be mapped onto Eq. (\ref{spinHam}), i.e. onto a Heisenberg model when the calculated exchange parameters are spin configuration-independent. Note that in the exact collinear limit (e.g. in ferromagnetic ground state) Eq. (\ref{Jncol}) reduces to the expression $A_{ij}^{00}-A_{ij}^{zz}$, and it can be shown that this is equal to the LKAG formula given by $A_{ij}^{\uparrow \downarrow}$.

\subsection{Element and site resolved damping parameters}\label{sec_sergey}


Within the present work, the Gilbert damping  parameter
 is calculated on the basis of  the
linear response formalism    \cite{EMKK11}.
The approach used derives from a representation of the electronic structure in terms
of  the Green functions $G^{+}(E)$ that in turn is determined
by means of  the multiple scattering formalism \cite{ebert}.
The diagonal elements $\mu=x,y,z$ of the  Gilbert damping
tensor can be written as  \cite{EMKK11}:
%
{\begin{eqnarray}
\alpha^{\mu\mu} =   \frac{g}{\pi m_{tot}} \sum_{j } \mbox{ Tr}
\left\langle \underline{\mathcal{T}}^{\mu}_0 \,
 \tilde{\underline{\tau}}_{0j}\,
\underline{\mathcal{T}}^{\mu}_{j} \,
 \tilde{\underline{\tau}}_{j0} \right\rangle_{c} \; ,
\label{alpha_MST}
\end{eqnarray}
where the effective g-factor  $g = 2(1 + {m_{orb}}/{m_{spin}})$
and  total magnetic moment
 $ m_{tot} = m_{spin} + m_{orb} $ are given
by the   spin and orbital moments, $m_{spin}$ and $m_{orb}$,
respectively, ascribed to a unit cell. Eq.~\ (\ref{alpha_MST}) gives $\alpha^{\mu\mu}$ for the atomic cell at lattice site $0$
and implies a summation over contributions from all sites indexed by $j$
including $j=0$.
The elements of the matrix $\tilde{\underline{\tau}}_{0j}$ are given by
 $\tilde{\tau}^{\Lambda\Lambda'}_{0j} =
\frac{1}{2i}( \tau^{\Lambda\Lambda'}_{0j} - \tau^{\Lambda'\Lambda}_{0j})$
where {$\underline{\tau}_{0j}$ is the
so-called  SPO matrix  \cite{EMKK11} evaluated for the  Fermi energy, $E_F$.
Finally, the {matrix} $\underline{\mathcal{T}}^{\mu}_j$}
is represented by the matrix elements
%
{\begin{eqnarray}
\mathcal{T}^{\mu, \Lambda'\Lambda}_{j}
& = & \int d^3r\; (Z_{j}^{\Lambda'}(\vec{r}))^{\times}\;\left[\beta
\sigma^{\mu}B_{xc}(\vec{r})\right] Z_{j}^{\Lambda}(\vec{r})
\label{matrix-element}
\; ,
\end{eqnarray}
%
of the {torque operator $\hat{\mathcal{T}}^{\mu} = \beta(\vec{\sigma}\times \hat{m}_{z})_{\mu}B_{xc}(\vec{r})$ \cite{EM09a}}.
Here,   $Z_{j}^{\Lambda}(\vec r)$ is a regular solution to the single-site
Dirac equation
for the  Fermi energy $E_F$
labeled by the combined quantum numbers $\Lambda$
($\Lambda = (\kappa,\mu)$), with $\kappa$ and $\mu$  being the
spin-orbit and magnetic quantum numbers \cite{Ros61}.

To calculate the configurational average
indicated by the brackets $\langle ... \rangle_c$, in the case of disordered
alloys, the CPA alloy theory is used.
This is done using the scheme developed by Butler \cite{But85}
in the context of electrical conductivity,
that splits the {summation in Eq.\ (\ref{alpha_MST})
into a site diagonal part,
$\left\langle \underline{\mathcal{T}}_{0}{\mu} \,
 \tilde{\underline{\tau}}_{00}\,
\underline{\mathcal{T}}_{0}^{\mu} \,
 \tilde{\underline{\tau}}_{00} \right\rangle_{c} $,
and a site off-diagonal part,
$ \sum_{j\ne0 }
\left\langle \underline{\mathcal{T}}_{0}^{\mu} \,
 \tilde{\underline{\tau}}_{0j}\,
\underline{\mathcal{T}}_{j}^{\mu} \,
 \tilde{\underline{\tau}}_{j0}  \right\rangle_{c} $,
respectively}.
Dealing with the second term one has to
account in particular for the so-called ``in-scattering processes" that deals with vertex corrections of crucial importance for the Gilbert damping \cite{MKWE13}.

As indicated above,   Eq.\ (\ref{alpha_MST}) gives in the case
of a unit cell involving in the case of an alloy several  atomic types
a value for  $\alpha^{\mu\mu}$ that is averaged  over these types.
In the case of a system consisting only of magnetic components,
i.e.\ none of its components has an induced magnetic moment,
one may also introduce a type-projected
damping parameter $\alpha_{t}^{\mu\mu}$. As the average
for the site diagonal as well as  site off-diagonal
contributions to  $\alpha^{\mu\mu}$ involve a
sum over  the types $t$ with the type-specific
contribution weighted by the corresponding concentration  $x_t$  \cite{But85}
one is led in a natural way to the expression:
%
{\begin{eqnarray}
\alpha_{t}^{\mu\mu} &=&  \frac{g_t}{\pi m^{t}} \,{\rm Tr }\, \underline{\mathcal{T}}_{0}^{\mu}  \Big[
 \Big<
\underline{\tau}_{00}
\underline{\mathcal{T}}_{0}^{\mu}\underline{\tau}_{00}
\Big>_{t \, on \, 0} \nonumber \\
&& + \sum_{j \neq 0} \sum_{t' \, on \, j} x_{t'} \,\Big<
\tilde{\underline{\tau}}_{0j}
\underline{\mathcal{T}}_{j}^{\mu}\tilde{\underline{\tau}}_{j0} \Big>_{t \, on \,0; \, t' \, on \, j}\Big]
\;
\label{alpha_MST2}
\end{eqnarray}
%
with $t$ and $t'$ denoting the atomic types at the lattice positions $0$ and
$j$, respectively.
Here, we  use a  type-specific  g-factor $g_t$ and
magnetic moment $m^{t}$ given by the corresponding
 spin and orbital moments, $m_{spin}^t$ and $m_{orb}^t$, respectively.
The resulting  definition for the element projected Gilbert damping $\alpha_{t}^{\mu\mu}$
leads now to an average for the unit cell according to:
$\alpha^{\mu\mu} = \sum_t x_t \alpha_{t}^{\mu\mu} $.
 Because of the  normalizing factor ${g}/{m_{tot}}$ used
in   Eq.\ (\ref{alpha_MST}) this expression will lead in general
to results slightly deviating from that based on  Eq.\ (\ref{alpha_MST}).

The calculations of the Gilbert damping parameter for
 finite temperature presented below have been done  using the so
called alloy analogy model \cite{EMKK11}. This approach is based on
 the adiabatic approximation
assuming  random temperature dependent
displacements of the atoms from their equilibrium
positions. Using a discrete set of displacements with each
displacement treated as an alloy component, the problem of
calculating the  thermal average for a given temperature $T $
is reduced to the problem of calculating the
configurational average as done
for substitutional alloys \cite{EMKK11}.

\subsection{Atomistic spin dynamics}
\par
The temperature dependent evolution of spins are calculated from atomistic spin-dynamics (ASD) simulation at different temperatures using the framework of Landau-Lifshitz-Gilbert (LLG) formalism.
The temporal evolution of an atomic moment in LLG formalism is given by\cite{asd1},
{\begin{eqnarray}
\label{llge}
\frac{d{\vec{m}}_{i}(t)}{dt} &=& -\frac{\gamma}{(1+\alpha^2)} \left( { \vec{ m}}_{i}(t)\times \vec{B}_{i}(t)+ \right. \\ \nonumber && \left. \frac{\alpha}{{\textit m_i}}
{\vec{m}}_{i}(t)\times ({\vec{m}}_{i}(t)\times \vec{ B}_{i}(t)) \right),
\end{eqnarray}
where $\gamma$ is the gyromagnetic ratio, $\alpha$ represents the dimensionless Gilbert damping constant and ${\vec{m}}_i$ stands for an individual atomic moment on site $i$. Note that ${\vec{m}}_i= m_{i} \vec{e}_{i}$ where $m_{i}$ is the magnitude of the magnetic moment (at site $i$). The effective magnetic field is represented by $\vec{B}_{i}=-\frac{\partial {\cal H} }{\partial \vec{m}_{i}}+{\vec{b} }_i$, where $\mathcal{H}$ is given by Eq.
\eqref{spinHam2} 
and $\vec{b}_{i}$ 
 is a time evolved stochastic magnetic
field which depends on the spin temperature from the two-temperature (2T) model\cite{boven}.

The analytical expression of the two temperature model reads as,
\begin{eqnarray}
\label{tt}
T_{s}  & = & T_{0} + \\  \nonumber && (T_{P}-T_{0})\times(1- \exp^{(-t/\tau_{\rm initial})})\times \exp^{(-t/\tau_{\rm final})} + \\  \nonumber && (T_{F}-T_{0})\times(1-\exp^{(-t/\tau_{\rm final})})
\end{eqnarray}
where $T_{\rm s}$ is the spin temperature, $T_{\rm 0}$ is initial temperature of the system, $T_{\rm P}$ is the peak temperature after the laser pulse is applied and $T_{\rm F}$ is the final temperature. $\tau_{\rm initial}$ and $\tau_{\rm final}$ are exponential parameters.
The calculated spin temperature from Eq.~\eqref{tt}, is explicitly passed into LLG equation via the {stochastic magnetic field $\vec{b}_{i}$ in Eq.~\eqref{llge}, which takes into account  thermal fluctuations of the system and the strength of the stochastic field is defined as $D = \frac{\alpha k_{B}T_{s}}{\gamma m} $, $k_{\rm B}$ is the Boltzmann constant. Alloying Co is treated by spatial random disorder of the Mn dopant.

The dynamical structure factor, which describes the magnon dispersion relation, is obtained from the Fourier transform of space and time displaced correlation function

\begin{equation}
C^{\mu}({{\bf r},{\bf r'}},t)= \langle m^{\mu}_{{\bf r}}(t) m^{\mu}_{{\bf r'}} (0) \rangle -  \langle m^{\mu}_{{\bf r}}(t)\rangle \langle m^{\mu}_{{\bf r'}} (0) \rangle,
\end{equation}
where the ensemble average is represented in the angular brackets and $\mu=x,y,z$ is the Cartesian component, and its Fourier transform is written as,

\begin{equation}
S^{\mu}({\bf q} ,\omega) = \frac{1}{\sqrt{2\pi} N} \sum_{{\bf r}, {\bf r'}} e^{\mathrm{i}{\bf q}\cdot({\bf r}-{\bf r'})}\int^{\infty}_{-\infty} dt\;  e^{i \omega t} C^{\mu}({\bf r} , {\bf r'}, t),
\end{equation}
where {\bf q} and $\omega$ are the momentum and energy transfer,
respectively. $N$ is the number of terms in the summation.
\par To estimate the Curie temperatures we used the fourth order size dependent Binder cumulant\amend{~\cite{binder}}, which is defined as,
\begin{equation}
{U}_{L}=1 - \frac{{\langle  M^{4} \rangle}_{L} } { 3 {\langle M^{2} \rangle}^{2}_{L}},
\end{equation}
where $M$ is the total or average magnetization. $\langle \hdots \rangle$ is the ensemble and time average. Binder cumulants exploits the critical point and critical exponents in a phase transition from the crossing point of magnetization curves for different sizes $L$ of the system.

\section{Numerical details}\label{numerical}

The Perdew, Burke and Ernzerhof (PBE)\cite{gga} version of the generalized gradient approximation is used to describe the exchange-correlation potential. The spin polarized scalar relativistic full-potential (SR-FP) mode\cite{sprkkr} is used to calculate the total energies as a function of volume ($E(V)$) and the total ($M$) and element resolved ($m_{\rm Co}$, $m_{\rm Mn}$) magnetic moments. For the exchange integral ($J_{ij}$) and damping parameter ($\alpha$, $\alpha_{\rm Co}$, $\alpha_{\rm Mn}$) calculations, the potential is described within the atomic sphere approximation (ASA) using the scalar-relativistic (SR) and relativistic (R) mode, respectively \cite{sprkkr,MKWE13}. The basis set consisted of $s$, $p$, $d$ and $f$ orbitals ($l_{\rm max}$=3). The number of \textbf{k} points was set to $\approx$300, $\approx$500 and $\approx$1000000 for the calculation of the ground state properties, density of states (magnetic exchange integrals) and site and element resolved damping parameters, respectively. Equilibrium lattice constants are obtained by fitting $E(V)$ curves with a Morse type of equation of state\cite{morse}. The exchange constants are calculated up to 12 nearest neighbour shells.

We performed ASD simulations by using the UppASD software\cite{skubic,asd2} for the Co-based systems with a size of 20 $\times$ 20 $\times$ 20 unit cells and also with periodic boundary conditions. Here we used calculated exchange constants and averaging over 16 ensembles.

\section{RESULTS AND DISCUSSION}\label{result}

\subsection{Static properties of Co and Co-Mn alloys}\label{static}

\subsubsection{Electronic structure and magnetic properties}\label{electronic}
\begin{table*}[htp]

\centering 
\caption{Theoretically estimated lattice parameters ($a$), total ($M$) and element ($m_{\rm Co}$, $m_{\rm Mn}$) projected magnetic moments, Curie temperatures ($T_{\rm C}$), total and element projected densities of states at the Fermi level (DOS($E_{\rm F}$)) and Gilbert damping parameters ($\alpha$) calculated for bcc, fcc and hcp Co as well as for Co$_{1-x}$Mn$_{x}$ ($x = 0.1, 0.15, 0.2, 0.3$) alloys in the bcc and bct crystallographic phase. The experimental Curie temperatures, lattice constant, magnetic moment are shown in brackets and the DLM results are shown in parentheses. \amend{The magnetic moments are calculated using experimental lattice parameters.}}
\scalebox{0.9}{
\begin{tabular}{l| c| c c c| c| c c c|ccc }
\hline\hline
System &$a$/$c$ (\AA)  &\multicolumn{3}{c|}{$M$ ($\mu_{\rm B}$/atom)}& $T_{\rm C}$(K) &\multicolumn{3}{c|}{DOS($E_{\rm F}$) (states/Ry)}& \multicolumn{3}{c}{$\alpha$}\\
\hline
bcc Co                                 &2.85[2.83\cite{prinz}]&\multicolumn{3}{c|}{1.70[1.77\cite{diaz},1.50\cite{prinz1}]}&1280& \multicolumn{3}{c|}{25.3\amend{(31.448)}}&\multicolumn{3}{c}{0.0091\amend{(0.011)}} \\
fcc Co                                 &3.58[3.54\cite{fccco}]&\multicolumn{3}{c|}{1.62[1.68]\cite{liu}}&1311[1392\cite{vega}]&\multicolumn{3}{c|}{16.8\amend{(29.170)}}& \multicolumn{3}{c}{0.0057\amend{(0.009)}} \\
hcp Co                                &2.48/4.04[2.50/4.05]\cite{singal}&\multicolumn{3}{c|}{1.59[1.52]\cite{stearns}}&1306[1388\cite{crangle}]&\multicolumn{3}{c|}{12.8\amend{(28.68)}} & \multicolumn{3}{c}{0.0030\amend{(0.019)}}\\
\hline \hline
System     & & $M$ & $m_{\rm Co}$ &$m_{\rm Mn}$&~~~~~~& total&Co&Mn& $\alpha$ & $\alpha_{\rm 1}({\rm Co})$ & $\alpha_{\rm 2}({\rm Mn})$\\
\hline
Co$_{0.90}$Mn$_{0.10}$  \cite{acet}&2.86        &1.89&1.77(1.51)&2.99(2.75)&1280(1054)& 20.1& 21.6&  7.1 &0.0072&0.0083&0.0013\\
Co$_{0.85}$Mn$_{0.15}$  \cite{acet}&2.87        &1.96&1.78&2.98&1248& 19.5& 21.2&  9.1 &0.0066&0.0081&0.0015\\
Co$_{0.80}$Mn$_{0.20}$                 &2.88        &2.03&1.79&2.97&1129& 18.3& 20.2& 11.0 &0.0058&0.0076&0.0016\\
Co$_{0.70}$Mn$_{0.30}$                 &2.89        &2.13&1.80(1.46)&2.89(2.72)&1050(833)& 16.0& 17.5& 12.7 &0.0045&0.0061&0.0022\\
Co$_{0.90}$Mn$_{0.10}$ bct             &2.83 [2.71\cite{heiman}] &1.79&1.69&2.73& 1235[1215\cite{heiman}]   & 19.1& 20.2&  9.0 &0.0080&0.0090&0.0025\\
Co$_{0.70}$Mn$_{0.30}$ bct             &2.87 [2.90\cite{heiman}]  &2.11&1.80&2.85&    1054 [842\cite{heiman}]  & 16.8& 17.6& 14.9 &0.0047&0.0062&0.0025\\                             \hline \hline
\end{tabular}}\label{table1}
\end{table*}

The estimated theoretical lattice parameters ($a$) are listed in Table~\ref{table1} for bcc, fcc, and  hcp Co as well as Co$_{1-x}$Mn$_x$ alloys in the bcc and bct crystallographic phase, calculated using the PBE exchange-correlation functional. The local density approximation (LDA) calculations underestimates the lattice parameter with about 2\% when compared to PBE.} The presented PBE values for pure Co are in good agreement with the experimental data found for bcc Co grown in GaAs surface (2.82 \AA) \cite{Idzerda1989} and for fcc Co/Cu film (3.54 \AA) \cite{Harp1993}, as well as with the results of the previous DFT simulations \cite{Guo2000}. The Co-Mn alloys can be grown on a GaAs surface in bcc \cite{Dong1998, Wu2001} or bct \cite{Zhang2005, Zhao1997} crystal structure. The theoretical lattice constant $a$ of bcc alloys increases with Mn addition, which is consistent with the larger atomic radius for Mn compared to Co. The estimated lattice constants for $x$=0.3 (see Table~\ref{table1}) and $x$=0.4 (2.89 \AA) is in line with the experimental lattice parameters data reported for $x$=0.32 (0.4) \cite{Dong1998, Wu2001} which is 2.9 (2.89) \AA. The in-plane lattice parameter for the bct phase of Co$_{0.90}$Mn$_{0.10}$ and Co$_{0.70}$Mn$_{0.30}$ alloys are taken from experiments, Refs.~[\onlinecite{Zhang2005}] and [\onlinecite{Zhao1997}], respectively, while the out-of-plane lattice parameters have been optimised theoretically (see Table ~\ref{table1}).

Calculated densities of states (DOS) for pure Co in bcc, fcc and hcp crystal structure are presented in Fig. \ref{dos}. For these crystal structures, the 3$d$ majority spin channel is fully occupied, resulting in a low DOS at the Fermi level DOS$^{\uparrow}(E_{\rm F}$), while DOS$^{\downarrow}(E_{\rm F}$) lies near a peak in the 3$d$ DOS. The energy split between the majority and minority channels leads to a magnetic moment of 1.73$\mu_{\rm B}$ for the bcc lattice in good agreement with the previous theoretical data  \cite{Guo2000,diaz}. The experimental value for the magnetic moments of Co in the bcc structure are estimated from Co films grown on GaAs \cite{prinz1,bland1}. The average value is given as 1.4 $\mu_{\rm B}$ but in the centre of the film ($\unit[50]{\AA}$) the estimated experimental value for the Co magnetic moment in the bcc structure is  $\sim$1.7$\mu_{\rm B}$ \cite{bland1} which is in good agreement with the theoretically estimated value. For fcc Co, the calculated magnetic moment is in agreement with the previously published theoretical value of 1.64 $\mu_{\rm B}$ \cite{olle1,Guo2000} and in decent agreement with the experimental value of 1.68 $\mu_{\rm B}$\cite{liu}. Finally the magnetic moment of hcp Co is in good agreement with the reported experimental\cite{stearns} and theoretical data\cite{Guo2000}.

In both bcc and bct phases of Co-Mn alloys the DOS$^{\uparrow}(E_{\rm F})$ of Co is small due to the full occupation of the 3$d$ majority channels, and the shift in the occupation of  the majority and minority channels results in a magnetic moment higher than that in pure bcc Co and it is found to increase with increased Mn content (see Table~\ref{table1}). The Mn 3$d$ majority band is not fully occupied, and the 3$d$ minority band contains less states than in case of Co, due to the reduced number of electrons for Mn. The exchange splitting results in a higher magnetic moment for Mn than for Co (see Table~\ref{table1}). As Table~\ref{table1} shows for Co-Mn alloys the coupling between Co and Mn moments is ferromagnetic, with a large moment on both atoms. The results for the bcc structure give larger moments compared to data for the bct structure. 

\begin{figure}[h]
\includegraphics[scale=0.35]{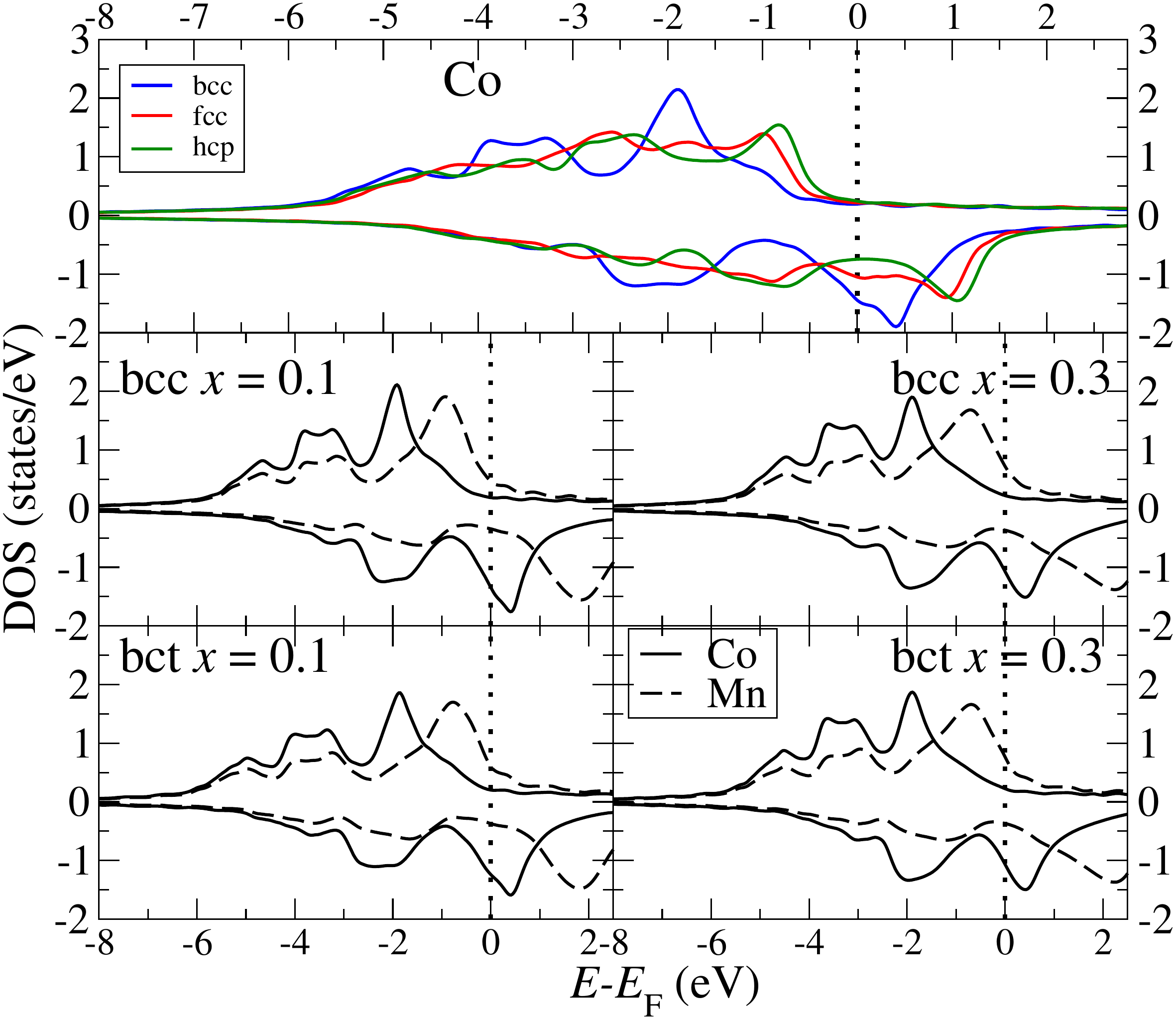}
\caption{(Color online) Density of states (DOS) per atom of bcc Co (blue), fcc Co (red), hcp Co (green) (upper panel) and bcc and bct Co$_{1-x}$Mn$_x$(lower panel). DOS for Co in Co$_{1-x}$Mn$_x$ alloys is labeled by full line while DOS of Mn is represented by dashed line. Dotted line illustrates the Fermi-energy.}
\label{dos}
\end{figure}

The smallest magnetic moments are given by the SR-FP mode. The SR-ASA (R spin) moments are in average 0.6 (0.5) \% larger compared to the SR-FP moments. We find the same trends for SR-ASA and R spin moments as a function of composition and structure as in the case of SR-FP moments. The orbital moment of Co is 0.085 $\mu_{\rm B}$ in the bcc phase. Its variation among different crystal structures and alloying is within 7\% and follows the same trend as for the spin moments. The orbital moment of Mn  in bcc Co$_{0.9}$Mn$_{0.1}$ is 0.018 $\mu_{\rm B}$. This value decrease to 0.016 $\mu_{\rm B}$ for $x$=0.3 Mn content in the bcc phase. The orbital moment of Mn in the bct phase is smaller compared to its value in the bcc phase for the corresponding composition.
 Local magnetic moments of Co and Mn in the DLM phase of Co$_{1-x}$Mn$_x$ for $x=0.1$ and $0.3$ are also presented in Table~\ref{table1}. Here we find that $m_{\rm Co}$ and $m_{\rm Mn}$ are reduced with 15\% and 20\%, respectively, in the DLM phase compared to that of FM solution.

All entries in Table~\ref{table1} show that the DLM configuration results in lower magnetic moments than for the FM configuration. To analyse this further we calculated the size of the magnetic moment of a supercell of 16 atoms in a bcc lattice, in which only the central atom had its magnetic moment rotated away from the z-axis with an angle $\theta$.
The rotated moment is denoted as $m_i$ and the rest of the spins are labeled $m_{j}$.
The self-consistently obtained values of  $m_i$ and  $m_{j}$ are shown for each value of $\theta$ in Fig.~\ref{yaroslav1}.
\begin{figure}[!h]
\centering
\includegraphics[angle=0,width=70mm]{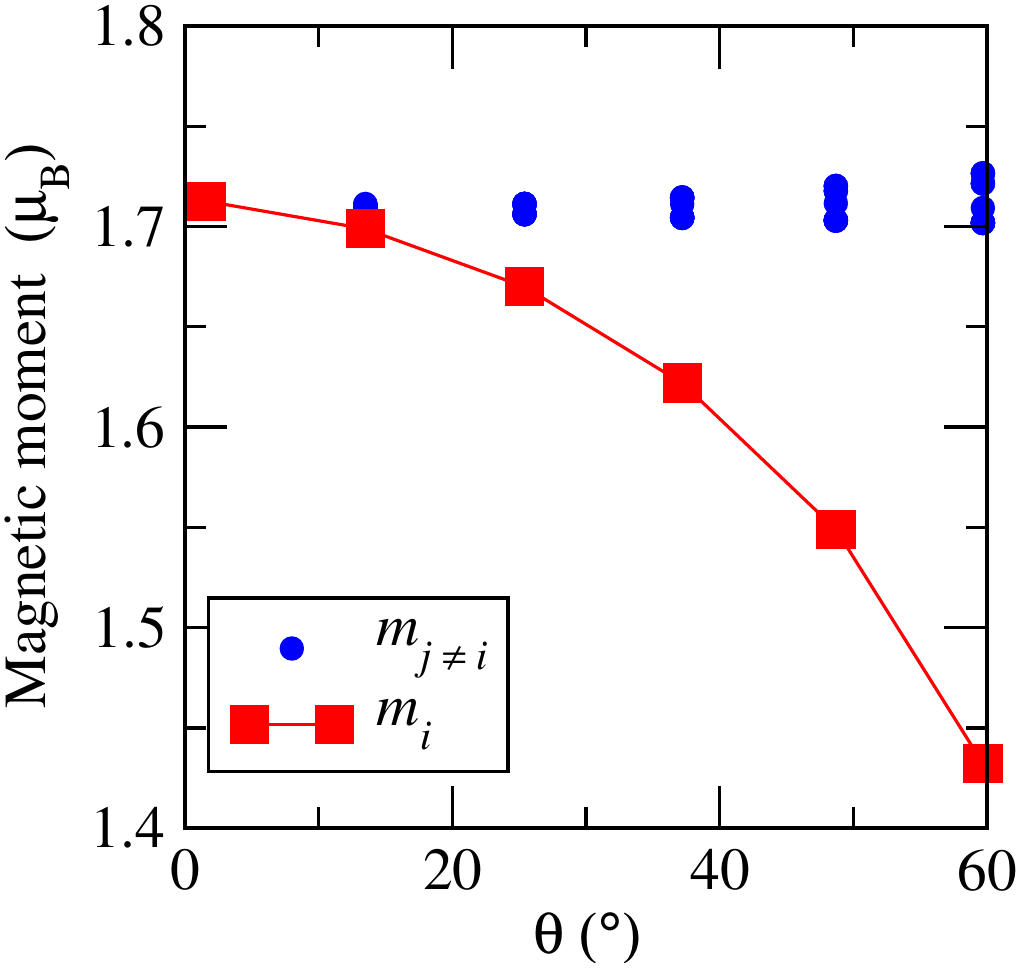}
\caption {\amendyk{Calculated magnetic moment as a function of rotation of a single spin with an angle $\theta$, in the FM background of bcc Co.}}
\label{yaroslav1}
\end{figure}
One can see that once $\theta$ increases, the magnitude of the moment $m_i$ tends to decrease.
We repeated the same calculations for bcc Fe and obtained qualitatively the same behaviour. 
Thus, the results of Fig.~\ref{yaroslav1} are consistent with the data in Table~\ref{table1}, and seem to reflect a quite general phenomenon that the rotation of a moment in a system with predominant FM interactions leads to the decrease of its length.
\amend{
This fact can be understood on the basis of a simple model, containing the energy of longitudinal spin variation (containing even powers of magnetization, as appropriate for a Landau expansion) and a nearest-neighbour exchange coupling ($J_1$).
In the case of the single spin rotation in the ferromagnetic background, one obtains:
\begin{eqnarray}
E = -\alpha_1 m^2_1 -\alpha_2 M^2 + \beta_1 m^4_1 + \beta_2 M^4 - \tilde{J}_1 \vec m_i \cdot \vec M, \quad 
\end{eqnarray}
where $M=\sum^N_{j=1} m_j$ represents the macro-spin formed by all $N$ nearest neighbour spins from the FM background.
The parameters $\alpha_1$, $\alpha_2$, $\beta_1$ and $\beta_2$ are phenomenological constants originating from the local exchange interactions.
The energy penalty stemming from the Heisenberg term when rotating the moment $m_i$ with an angle $\theta$ will be given by:
\begin{eqnarray}
E(\theta) - E(0) = \tilde{J}_1 m_i M (1-\cos(\theta)).
\end{eqnarray}
It is straightforward to show that, if the magnitude of $m_i$ is allowed to change, the system will try to minimize the energy costs of the single spin rotation by decreasing its length. In principle a reduction of  $M$ would also reduce the energy cost of rotating a single spin, but the Landau parameters describing this change are not in favor of this scenario. 
Finding the minima of the energy with respect to $m_i$ leads to the solution of non-linear equation, which can be solved numerically.
The numerical results confirm that with an increase of $\theta$, the value of $m_i$ corresponding to the minimum of the energy goes down.}

\begin{figure}[!h]
\centering
\includegraphics[angle=0,width=70mm]{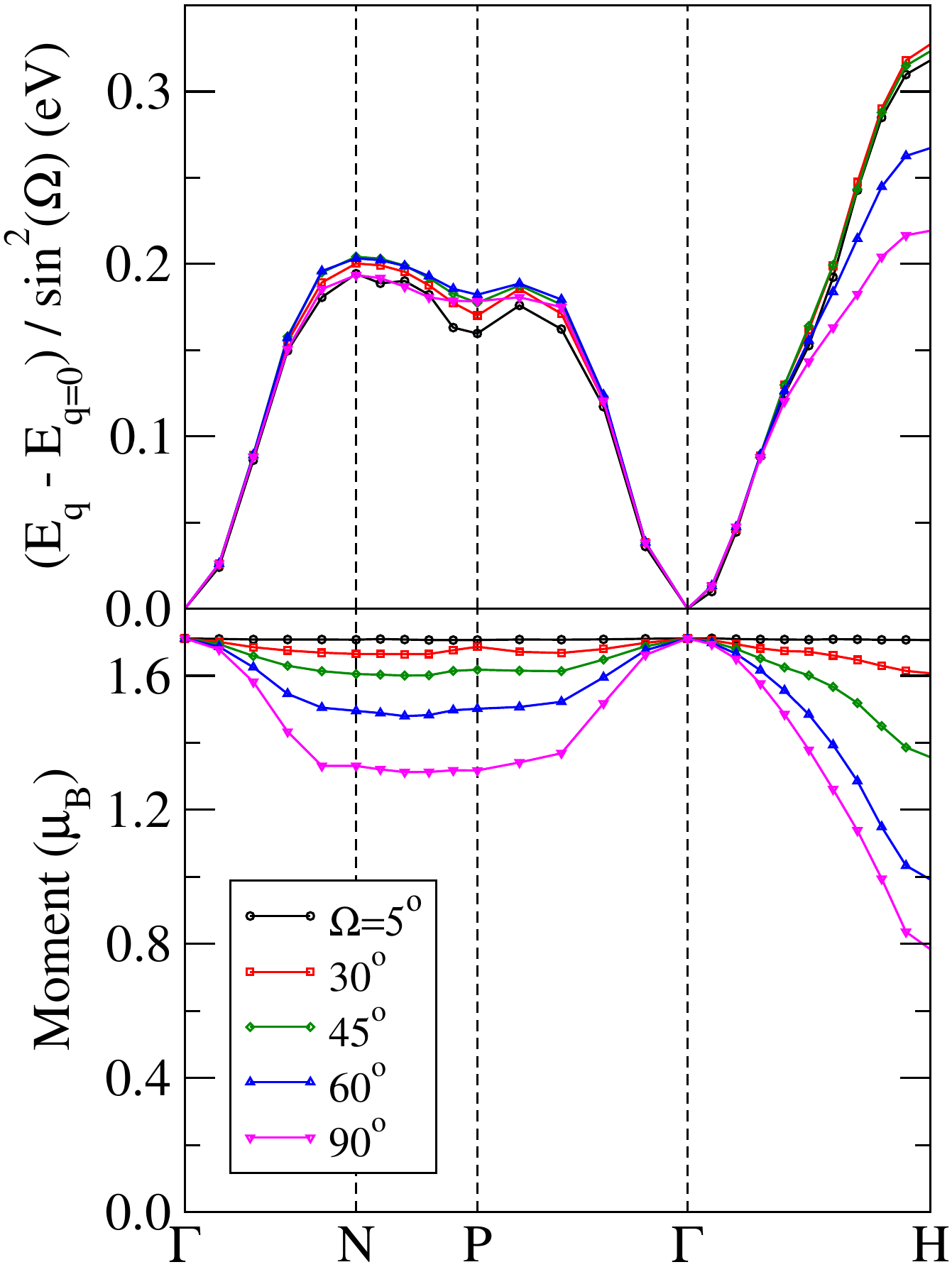}
\caption {\amendyk{Calculated relative spin spiral energies for different values of cone angle $\Omega$ along with the self-consistent values of the magnetic moment as a function of the propagation vector $\vec q$.}}
\label{yaroslav2}
\end{figure}
\amend{
Thus, for the case of a single spin rotation in bcc Co, we have shown that the magnitude of the magnetic moments unavoidably depends on the magnetic configuration.
In order to quantify how sensitive the magnetic excitations to inter-atomic non-collinearity, we have performed a series of spin spiral calculations\cite{Gen-Bloch} for the same structure. Spin spiral states are characterized by the propagation direction ($\vec q$) and the cone angle $\Omega$ between the magnetization and $\vec q$ vectors. Note that spin spirals with infinitesimal $\Omega$ would correspond to the actual magnon excitation. The bottom panel of Fig.~\ref{yaroslav2} shows the self-consistently obtained value of the magnetic moment in all different spin spiral states. Just as in the case of the single spin rotation (Fig.~\ref{yaroslav1}), the magnetic moment experiences a variation when $\vec q$ is changed. 
In the top panel of Fig.~\ref{yaroslav2} we show the relative energies of the spin spirals calculated for various $\vec q$ and $\Omega $ values. On $y$ axis we plot $E_{\vec q} - E_{q=0} / sin^2(\Omega)$, which is supposed to be $\Omega$-independent for a truly Heisenberg magnet (see e.g. Ref.~\onlinecite{halilov}). 
It is clearly seen from Fig.~\ref{yaroslav2} that, despite the changes in the magnetic moment values, all curves lie nearly on top of each other, if $\Omega$ lies within the range of 5 to 45 degrees. At larger $\Omega$ angles, most of spin spiral energies are still very close to each other and the largest differences appear for $\vec q$-vectors along $\Gamma$-H direction. Hence, one can see that Co is a remarkable system, which is characterized by a configuration-independent magnetic excitations in a wide range of magnetic states.
From Fig.~\ref{yaroslav2} we estimated that the Heisenberg Hamiltonian (Eq.~\eqref{spinHam}) is perfectly valid up to a critical value of the angle between the nearest-neighbouring spins of about 90 degrees. Quite importantly, the results indicate an intriguing interplay between the strength of the $\tilde{J}_{ij}$'s and the magnitude of $\vec m_i$'s, which tend to balance each other resulting in a configuration-independent $J_{ij}$'s.}

\begin{figure}[h]
\includegraphics[scale=0.35]{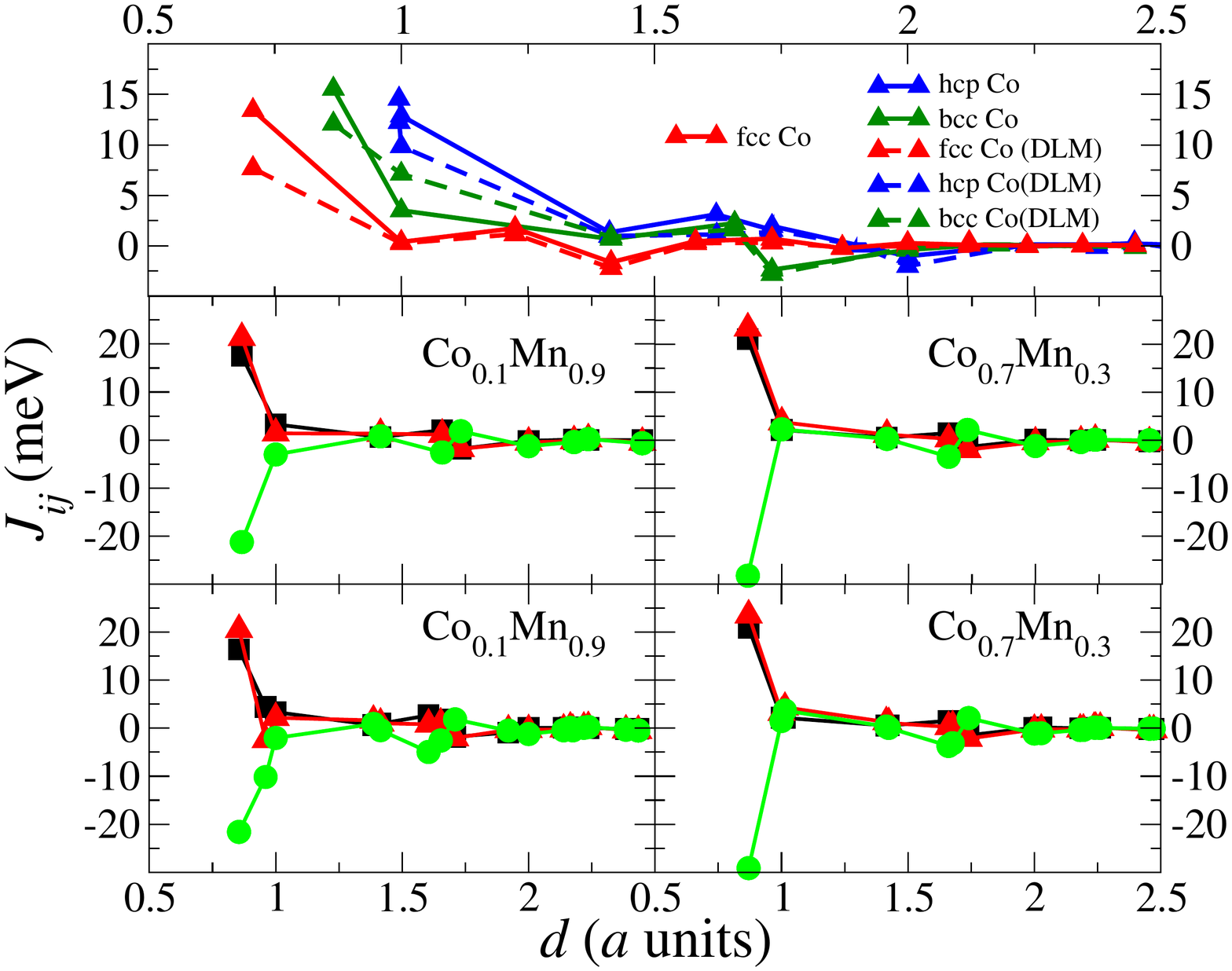}
\caption{(Color online) Exchange integrals ($J_{ij}$) from FM configuration for fcc, hcp and bcc Co (up triangle) are represented in solid lines, \amend{DLM (with dashed lines) and bcc} and bct Co$_{1-x}$Mn$_x$ alloys plotted as a function of distance $d$ (in $a$ lattice parameter units).  $J_{\rm Co-Co}$ are labeled by filled red triangles, $J_{\rm Co-Mn}$ by filled black squares and $J_{\rm Mn-Mn}$ by filled green circles. \amend{The exchange parameters are obtained for the FM state.}}
\label{jij}
\end{figure}

\subsubsection{Temperature dependent exchange interactions $J_{ij}$}\label{exchange}
The calculated exchange interactions, $J_{ij}$, for all Co-based systems at $T=\unit[0]{K}$ are plotted in Fig.~\ref{jij}. The Co-Co interactions have positive values showing a ferromagnetic coupling between Co atoms. In the Co-Mn alloys, the Mn atoms are ferromagneticaly coupled to the Co atoms and favour antiparallel coupling to the nearest Mn atoms, while the Co-Co interactions are ferromagnetic. All $J_{ij}$'s decay fast with distance. Increased Mn content is found to enhance all interactions, which can be explained by an increase of $m_{Co}$ and $m_{Mn}$, since according to Eq.~\eqref{spinHam}, the magnitude of the moments is effectively contained within the $J_{ij}$'s. The results obtained for the different phases of elemental Co are in overall good agreement with prior DFT studies, the differences coming from the employed computational methods.\cite{yarov1,yarov2,yarov3,yarov4,yarov5}

To continue the analysis of exchange interactions in these systems we investigated whether hcp, fcc, and bcc Co have Heisenberg exchange parameters that are configuration (temperature) dependent. To this end we determined $J^{nc}_{ij}$ defined by Eq.~\eqref{Jncol}. Note that the second term in Eq.~\eqref{gen} did not give a significant contribution in Co systems we considered here. We compare these results to those of bcc Fe which has been already shown to have $J_{ij}$'s that are configuration dependent, and hence not to be a perfect Heisenberg system\cite{yarov}. Note that $\vec{e}_{i}$ in Eq.~\eqref{gen} denotes the direction of the spin at site $i$, which can be formulated as $e_i=e(\theta_{i},\phi_{i})$ where $\theta_{i}$ and $\phi_{i}$ are the polar and azimuthal angles of the spin direction, respectively. The most simple non-collinear spin configuration may be the case when one spin in a ferromagnetic background is being rotated by a finite angle $\theta$. The dashed lines in Fig.~\ref{nc1} show the $J^{nc}_{ij}$'s for the nearest neighbour couplings in bcc Co and bcc Fe.
We find that bcc Fe is more configuration dependent than bcc Co, i.e., bcc Co is closer to a ``perfect" Heisenberg system. 
However, Fig.~\ref{yaroslav1} shows that this story is somewhat more complex, since for single site rotation the magnetic moment changes significantly with angle of rotation. 
\amend{As was already demonstrated for the case of spin spirals (Fig.~\ref{yaroslav2}), there is no contradiction, since $J_{ij}$ value is defined in such a way that it contains the magnetic moment value in itself. Thus, we witness once again that bcc Co reflects the physics intrinsic to an ideal Heisenberg magnet despite the sensitivity of its magnetic moment to the environment.}

\begin{figure}[h]
\begin{center}
\begin{tabular}
[c]{c}%
\includegraphics[width=80mm]{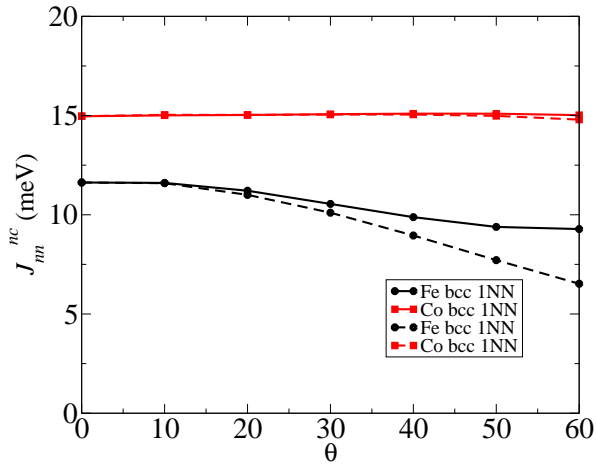}
\end{tabular}
\end{center}
\caption{(Color online) { Solid lines}: Non-collinear exchange coupling $J_{ij}^{nc}$ defined by Eq.~(\ref{Jncol}) for first neighbour spin pairs in bcc Fe and Co when one spin is fixed and the spin directions at its first neighbour sites are rotated by $\theta$ and $\phi$ .  { Dashed lines}:  Non-collinear exchange coupling $J_{ij}^{nc}$ defined by Eq.~(\ref{Jncol}) for first neighbour spin pairs in bcc Fe and Co when one spin is rotated by $\theta$ and $\phi$ . The azimuthal angle $\phi$ is set by a random number generator.}
\label{nc1}%
\end{figure}

\begin{figure}[h]
\includegraphics[scale=0.32]{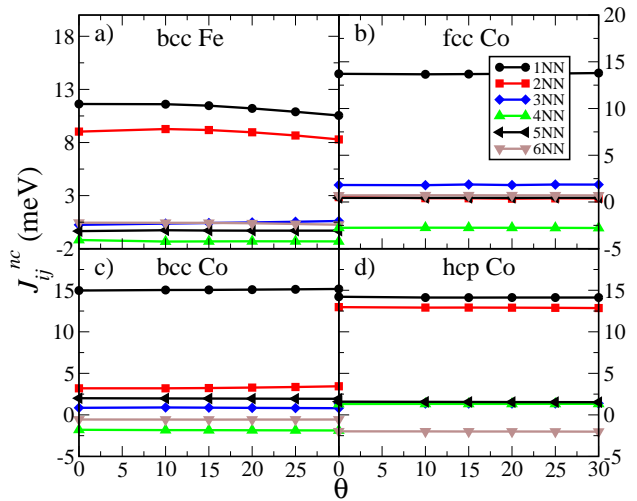}
\caption{(Color online) Non-collinear exchange coupling $J_{ij}^{nc}$ defined by Eq. (\ref{Jncol}) for the first six neighbour spin pairs in bcc Fe when one spin is fixed and the spin directions at its first neighbour sites are rotated by $\theta$ and $\phi$. Similar for the first six neighbour spin pairs in fcc, bcc and hcp Co systems receptively. In these figures, the azimuthal angle $\phi$ is set by a random number generator.}
\label{nc2}%
\end{figure}

A more realistic spin configuration can be constructed when one spin is fixed at, say, site $i$, and where the spin directions at its first neighbour sites are rotated by $\theta$ and $\phi$. These results are shown by the solid lines in Fig.~\ref{nc1}, modelling a finite temperature disordered background. For such a configuration we also calculated the $J^{nc}_{ij}$'s for neighbours, with varying distance. Fig.~\ref{nc2} shows all the interatomic exchange coupling parameters for the  first six nearest neighbour shells in bcc Fe, fcc Co, bcc Co and hcp Co, respectively. As  can be seen, all the Co phases seem to have an excitation spectrum that is close to an ideal Heisenberg system, but again it is due to a decrease of the individual moments and an increase of the Heisenberg exchange interaction, ${\tilde J}_{ij}$, as given by Eq.~\eqref{spinHam2}. 

\amend{The findings presented above} motivate that atomistic spin dynamics simulations can be made from Heisenberg exchange parameters from collinear ferromagnetic phases of bcc, fcc and hcp Co, if the definition of Eq.~\eqref{spinHam} is used for the energy excitations. As mentioned above, of all systems investigated here elemental Co stands out to be unique in this regard.

The $J_{ij}$ of Co-Mn alloys are evaluated only in the DLM phase (not shown) and only for alloy composition $x=\unit[0.1]{}$ and $\unit[0.3]{}$, since the aforementioned approach is cumbersome to apply for random alloys due to methodological reasons. The nearest neighbour $J_{\rm Co-Co}$, $J_{\rm Co-Mn}$ and $J_{\rm Mn-Mn}$ interactions are reduced compared to the value of the pure element by $21 \%$, $26 \%$, $7 \%$, respectively, for $x=\unit[0.1]{}$ and $35 \%$, $24 \%$, $38 \%$, respectively, for $x=\unit[0.3]{}$. This variations are in the same order as those obtained for Co in the DLM phase. Hence, Co and doped Co tend to behave like a "bad" Heiseberg system close to the phase transition temperature. In the low temperature regime, however, exchange parameter of Mn are likely to change (not shown here), similar to Fe, giving evidence for a non-Heisenberg behavior of CoMn even at low temperature.
\subsubsection{Curie temperatures}\label{temperature}
The calculated and experimental values of the Curie temperatures of Co-based systems are presented in Table~\ref{table1}. The calculated Curie temperature values of bcc, fcc and hcp Co are in a rather good agreement with experimental data. The measured Curie temperatures of ferromagnetic Co$_{1-x}$Mn$_{x}$  alloys \amend{decreases linearly} with an increase of concentration of Mn and becomes zero around 0.4. The calculated Curie temperature of Co$_{1-x}$Mn$_{x}$ alloys decrease linearly up to 30\% of Mn and the $T_{\rm C}$'s are somewhat overestimated when compared with experimental values. We address this discrepancy to slight variations of the Heisenberg parameter with Mn doping at the phase transition temperature. This finding is supported by the reduced $T_{\rm C}$ values calculated in DLM configuration for Co$_{0.9}$Mn$_{0.1}$ and Co$_{0.7}$Mn$_{0.3}$ (see Table~\ref{table1}). Since the difference in the phase transition temperature between the FM and the DLM phase are small compared to the maximal temperature obtained for the simulated pump-probe experiment, \amend{we make the approximation that a small amount of Mn in Co-Mn alloys will not change the temperature independence of the Heisenberg exchange parameters.} Hence, the extracted FM collinear parameters are used in simulated pump-probe experiments of bcc, fcc and Co, as well as bcc and bct Co-Mn alloys, as detailed below.

\subsubsection{Damping parameters}\label{damping}
Results for the Gilbert damping parameter
 of pure Co as well as for  Co$_{1-x}$Mn$_{x}$ alloys
calculated at $T = 324$ K are presented in Table~\ref{table1} together with the DOS($E_{\rm F}$). In the case of
pure bcc Co, this temperature
corresponds approximately to the minimum of the $\alpha(T)$ curve,
that indicates  the cross-over  of the contributions due to the intraband
(dominating at low temperature) and the interband (dominating at
high temperature) electron scattering events.\cite{GIS07}
In the case of Co$_{1-x}$Mn$_{x}$ alloys, on the other hand,
 the interband spin-flip
scattering events are responsible for magnetization dissipation in the
whole temperature regime, similar to the case of Cu impurities in Ni \cite{EMKK11}.
When the temperature increases above room temperature (not shown here),
the thermal lattice vibrations lead for
 Co as well as  Co$_{1-x}$Mn$_{x}$ alloys to an increase  of
 $\alpha(T)$.

As can be seen in  Table~\ref{table1}, an increase of the Mn concentration $x$ for bcc Co$_{1-x}$Mn$_{x}$ results in a decrease of the Gilbert damping, which correlates well with a decrease of the total DOS($E_{\rm F}$). For the atom-resolved damping parameters we also find \amend{its} correlation \amend{with} the value of the DOS($E_{\rm F}$)/atom.  With increasing Mn concentration in these alloys, the damping \amend{parameter of Mn spins} and DOS($E_{\rm F}$)/atom \amend{both tend to} increase, while the opposite is found for the Co \amendyk{spins}.
Comparing the Gilbert damping for the bcc and bct phases of Co$_{0.90}$Mn$_{0.10}$ (Co$_{0.70}$Mn$_{0.30}$), we find that the damping reaches higher values in bct phase than in bcc for both sublattices, which is in contrast to  the DOS$(E_{\rm F})$/atom. This point also indicates that other effects also play role in determining the damping parameter, and that the correlation between damping and DOS is strongest for alloys within the same crystal structure. 

\subsection{Dynamical properties}
\subsubsection{Dispersion relations}
The calculated dynamical structure factor or magnon spectrum $S^{z}({\bf q},\omega)$  of fcc, bcc and hcp Co along the high symmetry directions of the Brillouin zone are presented in Fig.~\ref{magnon}. These simulations used the exchange parameters reported in Fig.~\ref{jij} and the Gilbert damping of 0.005. The fcc Co magnon dispersion along $\Gamma $-X high-symmetry path is in good agreement with experimental magnon data\cite{pawel1}. The hcp Co magnon dispersion along $\Gamma $-M and $\Gamma$-A high-symmetry path agrees also quite well with experimental magnon data already reported in Refs.~[\onlinecite{shirane,perring}]. Also, the results in Fig.~\ref{magnon} are consistent with results published by Etz \etal \cite{etz-review}.


\begin{figure}[h]
\includegraphics[scale=0.5]{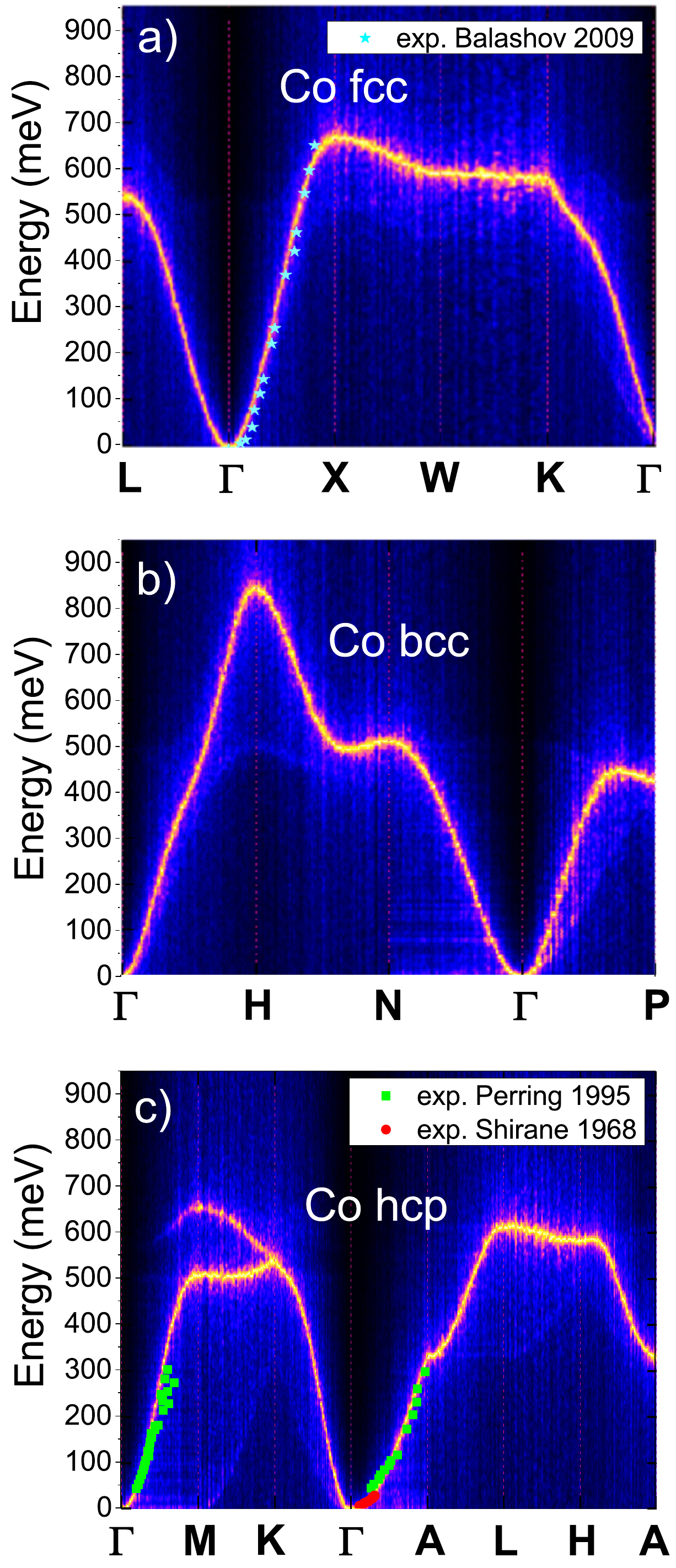}
\caption{(Color online) Magnon dispersion relation of fcc, bcc and hcp Cobalt along high symmetry directions of the Brillouin zone. Dots represent experimental measurements from Ref.~[\onlinecite{pawel1}]. \amend{All simulations were done assuming a negligibly low temperature and in FM state.}}
\label{magnon}
\end{figure}

\subsubsection{Ultrafast magnetization under laser fluence}
In the following section we present ultrafast magnetization dynamics of fcc, bcc and hcp Co as well as bcc and bct Co-Mn alloys under the influence of a femtosecond laser pulse. The results are obtained by the integration of the atomistic LLG equation in combination with the analytical two temperature model~\cite{raghugdfe}, and are plotted in Fig.~\ref{cobaltall}. The simulated laser pulse results in a temperature profile that initially starts at room temperature as initial temperature, $T_{\rm 0}$ and reaches its maximum at $T_{\rm P}$=1500K and finally relaxes to $T_{\rm F}$= 450K via several scattering processes (see top panel of Fig.~\ref{cobaltall}). \amend{The exponential parameters, $\tau_{\rm initial} = 1 \cdot 10^{-14}$ and $\tau_{\rm final} =3 \cdot 10^{-12}$ are used in the 2TM model.}
 Damping values taken from the first principles theory are considered in the simulations, called, $\alpha$, but also artificially enhanced values of the damping are used to investigate how the damping influence the ultrafast magnetization. The enhancement considered is 5 and 10 times larger, i.e. 5$\times\alpha$ and 10$\times\alpha$. The magnetization decreases rapidly to a minimum of about 44\% to 84\% of the total magnetization in fcc Co, 42\% to 84\% in bcc Co and 20\% to 57\% in hcp Co for different damping values, as shown in Fig.~\ref{cobaltall}.

 \begin{figure}[h]
\includegraphics[width=8.5cm,height=8.5cm]{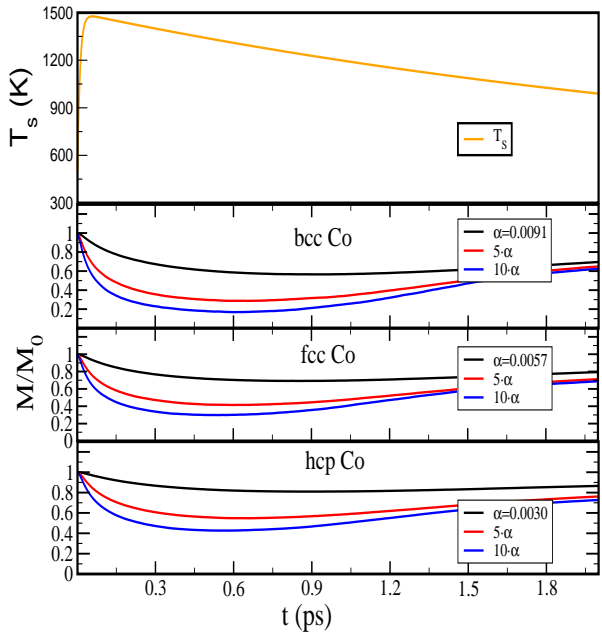}
\caption{(Color online)Time dependent normalized average magnetization (M/M$_{0}$) dynamics of Co systems after ultrashort laser irradiation (constant fluency) for different damping values \amend{with FM state}.~\amend{Top panel shows time dependent spin temperature.} \amend{The exchange parameters are obtained from SPR-KKR calculations.} Where M$_{0}$ is magentization at 300K }
\label{cobaltall}
\end{figure}

\begin{figure*}[!h]
\centering
\subfigure{\includegraphics[scale=0.3]{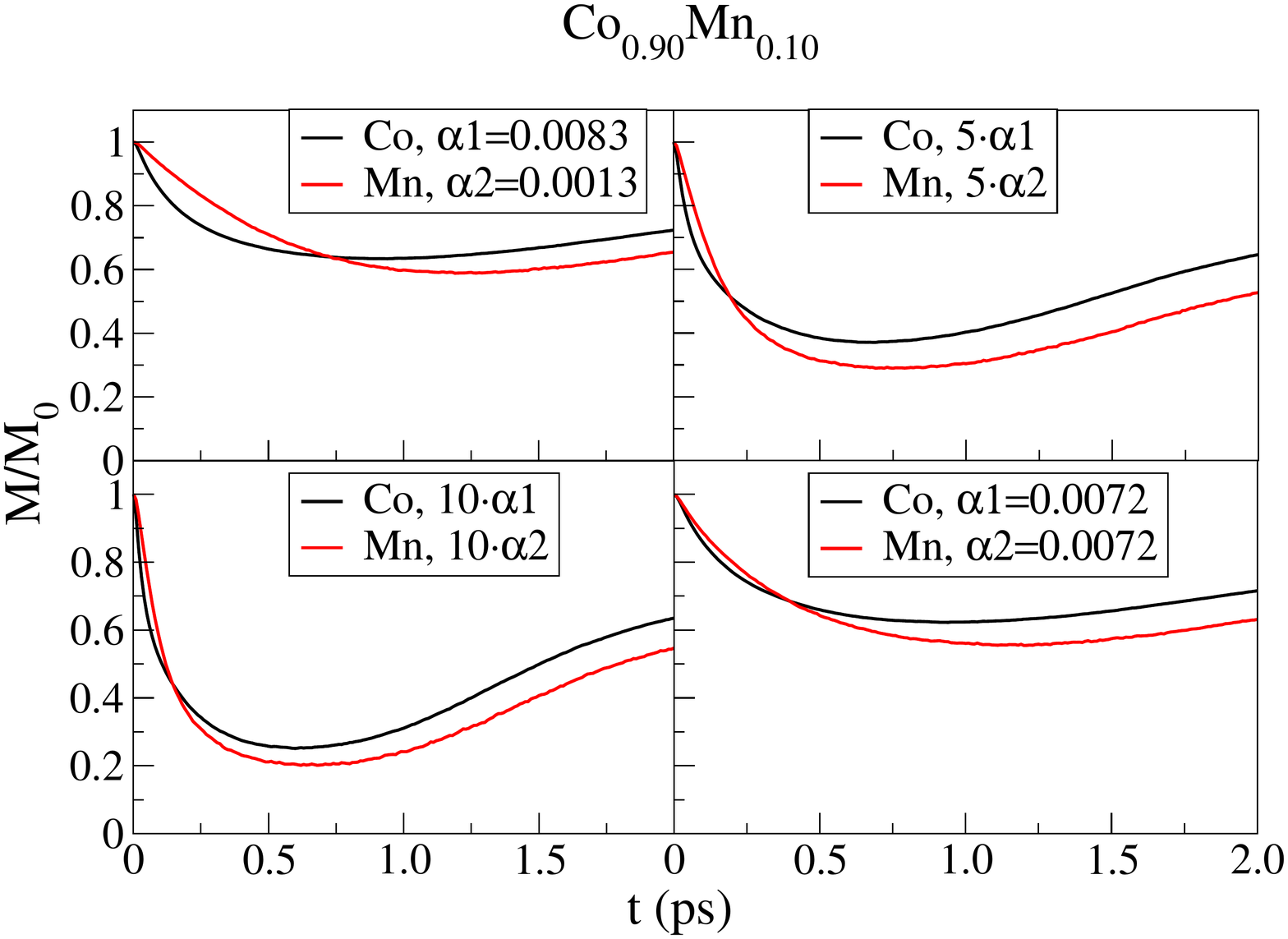}}
\subfigure{\includegraphics[scale=0.3]{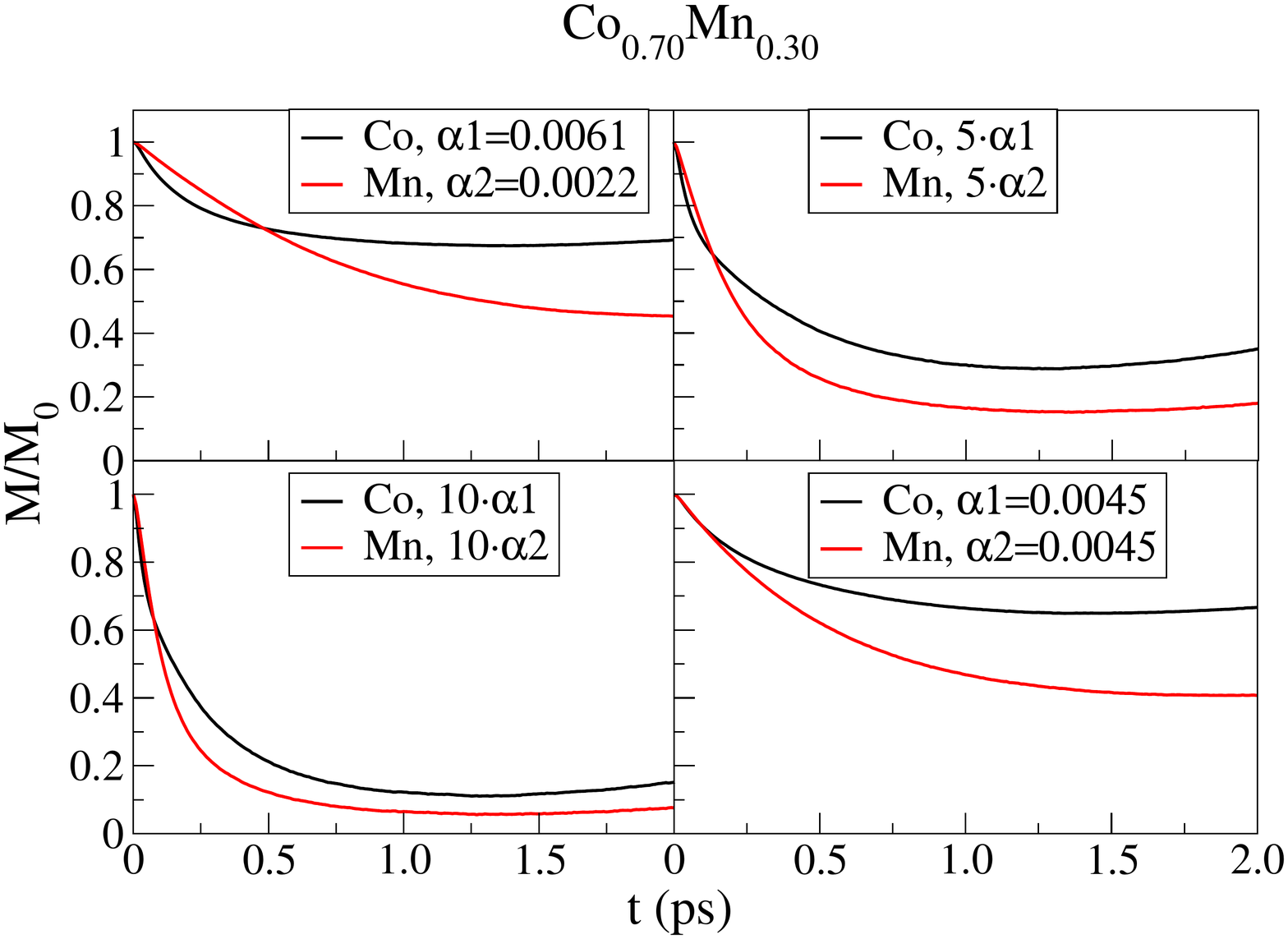}}
\subfigure{\includegraphics[scale=0.3]{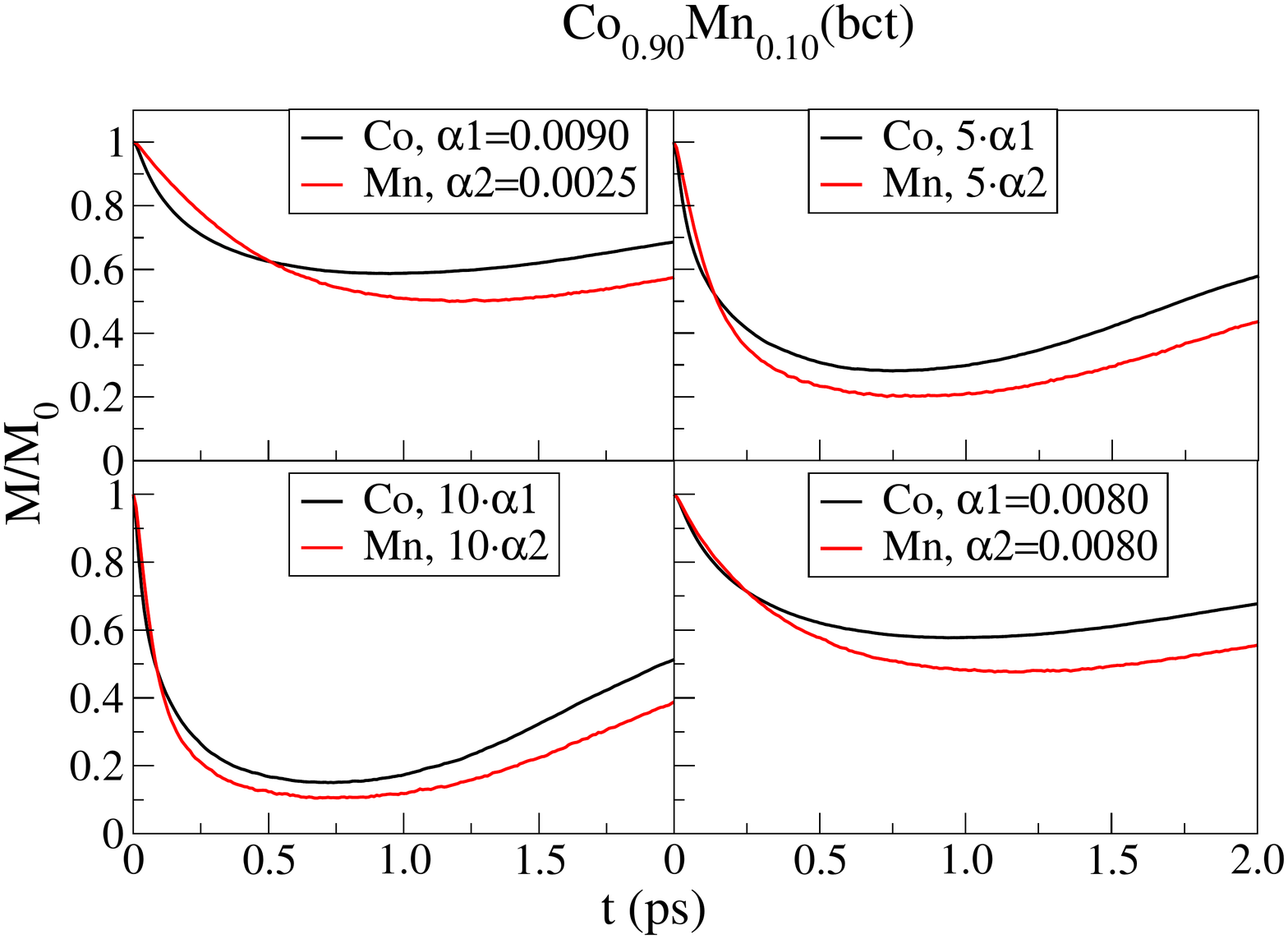}}
\subfigure{\includegraphics[scale=0.3]{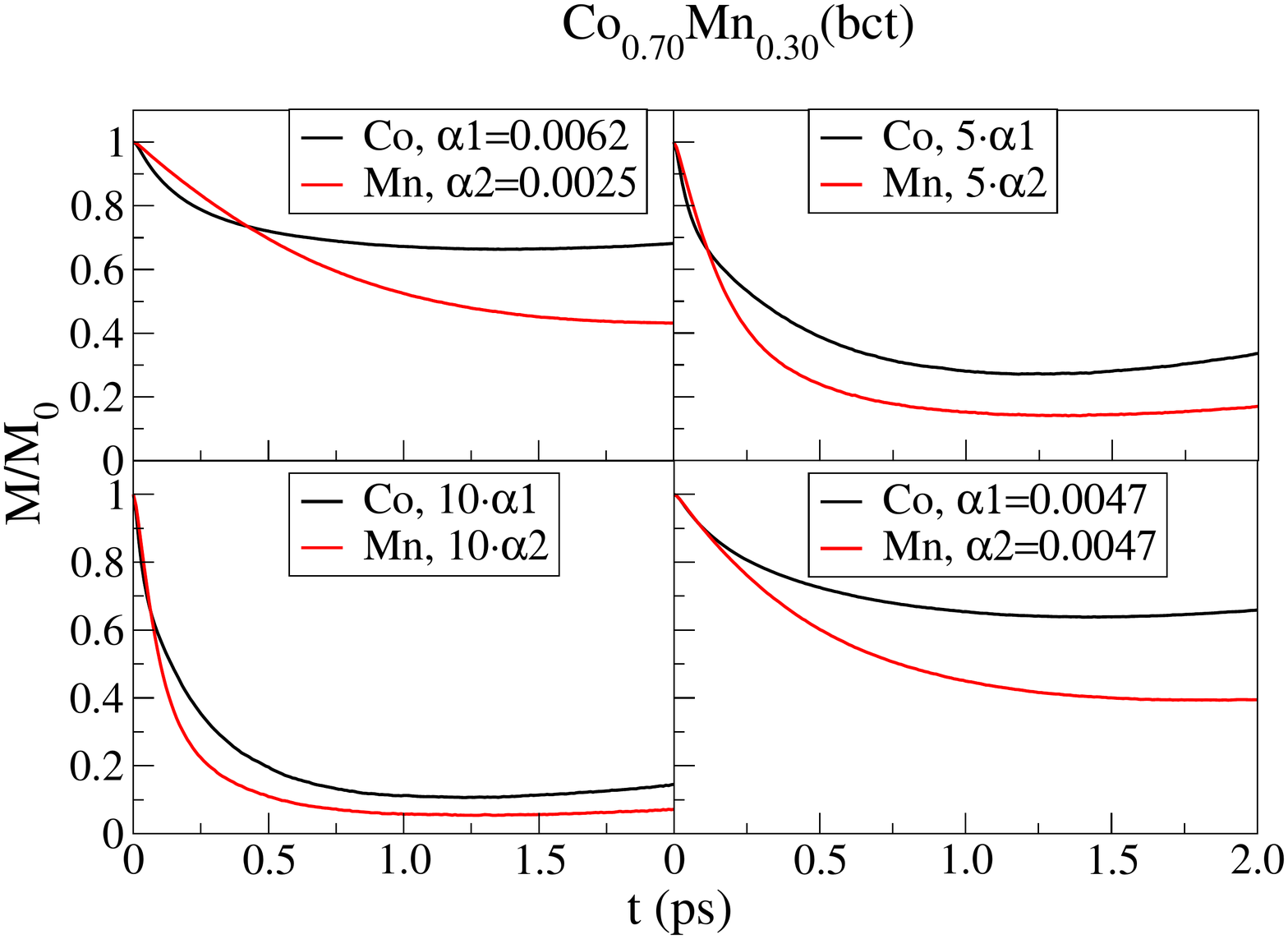}}
\color{white}
\color{black}
\caption{(Color online) Time evolution of the normalized average magnetization (M/M$_{0}$)
of Co$_{0.90}$Mn$_{0.10}$ and Co$_{0.85}$Mn$_{0.15}$ and alloys under
the influence of a thermal heat pulse for different sublattice damping parameters \amend{with FM state}. The black and red
lines represent the Co and Mn sublattices respectively. The damping, $\alpha1$ and $\alpha2$, refer to Co and Mn sublattices, \amendyk{respectively}. \amend{The exchange parameters are obtained from SPR-KKR calculations.}}
\label{fig3}
\end{figure*}

As Fig.~\ref{cobaltall} shows, the quenching of magnetization increases with the increase of damping parameter. This highlights the fact that both the demagnetisation time and the reduction of the magnetic moment in laser induced demagnetisation measurements depend critically upon the damping parameter. Furthermore, Fig.~\ref{fig3} shows that for systems with more than one magnetic sublattice, the magnetisation dynamics may be different and that these sublattices therefore display different demagnetisation times. The demagnetization times are calculated using double exponential fitting function,
as described in Ref.~[\onlinecite{mathias}] and $\tau_{m}$ are listed in Table~\ref{table2}. 
We would like to avoid a detailed comparison between the obtained theoretical and experimental demagnetization times because  the theoretical values are calculated for single crystal phases while the measurements are made for polycrystalline samples, potentially with several crystallographic phases present. However the gross features of the numbers listed in Table~\ref {table2} may be comparable to experimental data \cite{koopmans2}.
To end this section, in Table~\ref{table2}, it is shown that the demagnetization time is reduced for increasing value of the damping parameter in agreement with the finding published in Refs.~[\onlinecite{koopmans3,kantner}]. 

\begin{table}[htp]

\centering 
\caption{Element specific demagnetization times for Co-based systems. ${\alpha}_{\rm 1}$  refers to Co and  ${\alpha}_{\rm 2}$ refers to Mn. \amend{Experimental data from polycrystalline samples presented in brackets.}}
\scalebox{0.8}{\begin{tabular}{l| c c |c c }
\hline\hline
System     &  Co &  & Mn   \\
    &   $\tau_{\rm de}$(ps) & &   $\tau_{\rm de}$(ps)     \\

\hline
fcc Co ($\alpha= 0.0057$)  & 0.255 \amend{[0.15-0.25]}\cite{koopmans2}  &  &    \\
bcc Co ($\alpha= 0.0091$) & 0.250\amend{[0.15-0.25]}\cite{koopmans2}&  &      \\
hcp Co  ($\alpha= 0.0030$)  &0.300\amend{[0.15-0.25]}\cite{koopmans2} & \\
Co$_{0.90}$Mn$_{0.10}$ ($\alpha 1=0.0083,\alpha 2=0.0013$) &0.219&& 0.6    \\
Co$_{0.90}$Mn$_{0.10}$ ($\alpha=0.0072$) &0.22&& 0.375    \\
Co$_{0.70}$Mn$_{0.30}$  ($\alpha 1=0.0061,\alpha 2=0.0022$)             &0.23&& 1.0  \\
Co$_{0.70}$Mn$_{0.30}$  ($\alpha=0.0045$)             &0.5&& 0.65 \\
Co$_{0.90}$Mn$_{0.10}$ bct ($\alpha 1=0.0090,\alpha 2=0.0025$)         &0.215& & 0.56   \\
Co$_{0.90}$Mn$_{0.10}$ bct ($\alpha=0.0080$)         &0.225&& 0.375   \\
Co$_{0.70}$Mn$_{0.30}$ bct ($\alpha 1=0.0062,\alpha 2=0.0025$)        &0.22&& 0.7   \\
Co$_{0.70}$Mn$_{0.30}$ bct ($\alpha=0.0047$)        &0.45&& 0.6   \\                             \hline \hline
\end{tabular}}\label{table2}
\end{table}

\begin{figure}[h]
\includegraphics[width=8.5cm,height=8.5cm]{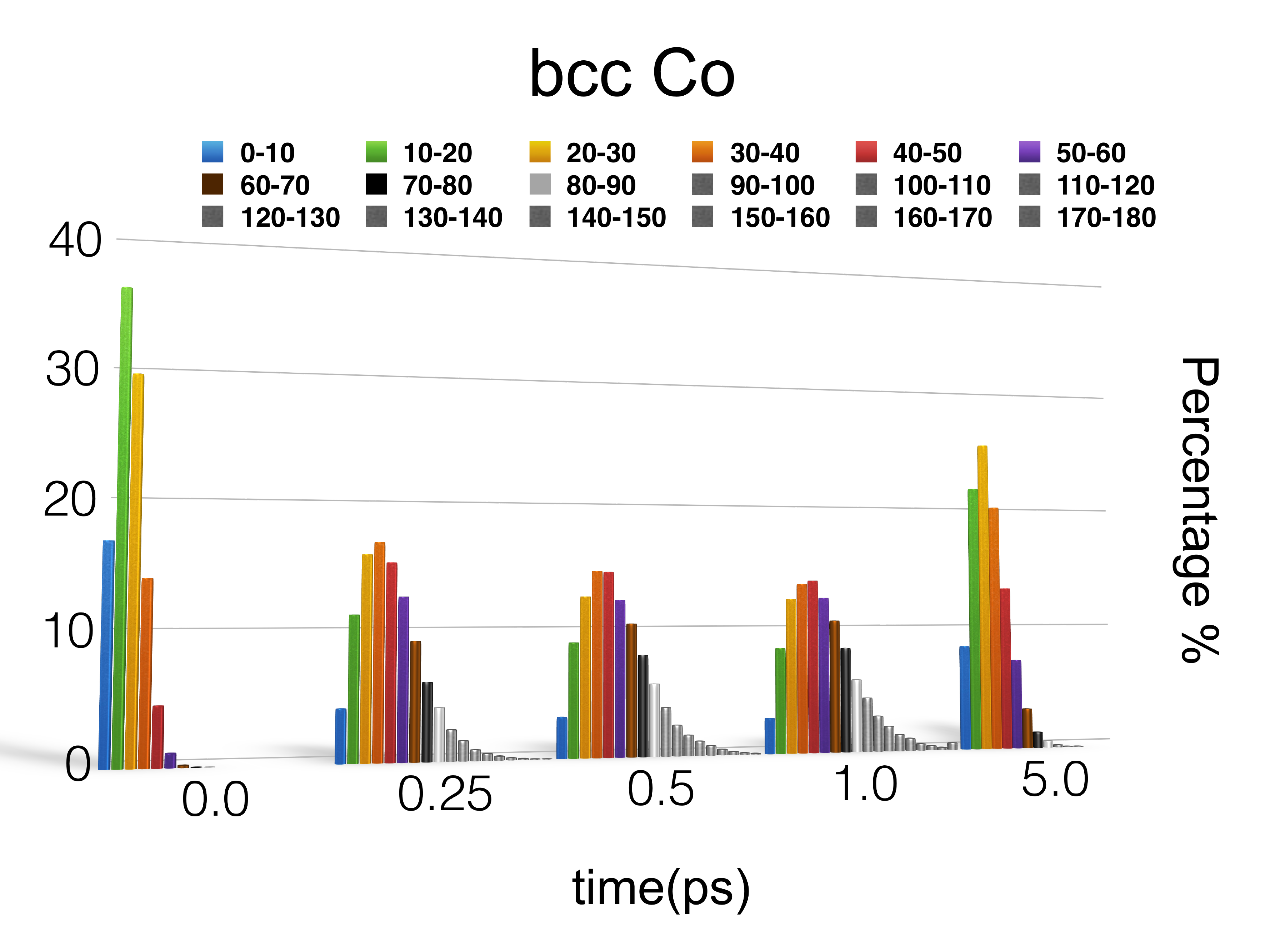}
\caption{(Color online) \amendyk{The time evolution of the distribution of the azimuthal angle $\theta$, for the different atomic spins for bcc Co. Note that the distribution is shown in intervals 0-10 degrees, 10-20 degrees etc. where each interval is shown as a bar with a specific colour. The distribution is shown for t=0 ps, 0.25 ps, 0.5 s, 1.0 ps and 5.0 ps after the laser pulse starts to heat up the sample. }}
\label{histo}
\end{figure}

\amend {In Fig. \ref{histo} we show the distribution of \amendyk{azimuthal angles} ($\theta$) of the atomic spins during the demagnetisation process, for bcc Co.
Note that at each time the distribution of the $\theta$ angles is found to follow essentially a Boltzmann distribution function.} This is not an obvious result since \amendyk{the atomic spins are in  out-of-equilibrium situation}. 
\amendyk{We have also calculated the angles between the nearest-neighboring spins (not shown) and found that at each time step these angles do not exceed 90 degrees.} 
\amendyk{As we have shown above (Fig.~\ref{yaroslav2}),} \amend{the magnetic excitation energies in bcc Co are represented accurately in this interval of angles using Eq.~\eqref{spinHam} (see also Fig.~\ref{nc1}), which lends credence to the approach adopted here to study ultrafast magnetisation dynamics.}

Next we analyze the Co-Mn alloys in more detail, and we focus on Co$_{0.90}$Mn$_{0.10}$ and Co$_{0.70}$Mn$_{0.30}$ in the bcc and bct structures, respectively. Element-specific damping parameters (${\alpha}_{\rm 1}$ (Co), ${\alpha}_{\rm 2}$ (Mn)) of Co-Mn alloys are used to investigate the angular momentum exchange between the sublattices in the spin dynamics simulations. Results are presented in Fig.~\ref{fig3}. The calculated element-specific de- and re-magnetization of Co-Mn alloys show a variety of possible situations that can be encountered in laser experiments on alloys with two magnetic sublattices. The demagnetization of Co precedes that of Mn by 0.15-0.6 ps due to the low damping value of Mn. The increase of damping parameters on both sublattices by 5 to 10 times in all considered alloys reduces the relative difference between the sublattice magnetism in the demagentization phase. Furthermore, our results show that the quenching of magnetization on both sublattices increases with the increased damping parameter. For the lowest damping value of Mn in bcc Co$_{0.70}$Mn$_{0.30}$, the Mn demagentization time is $\sim$4 times slower than that of the demagentization of Co. \amend{The results of Fig.~\ref{fig3} show that a large asymmetry in the damping parameter in multicomponent magnets is a good parameter to use when one wants to identify systems with very different behavior of the demagnetization in ultrafast pump probe experiments. We propose that this is a parameter that should be explored when one tries to identify alloys and compounds in which the element specific magnetisation dynamics is drastically 
different. We hope these results can motivate further experimental studies.}

\section{Conclusion}
\label{conclusion}
 With the extension of LKAG ab initio interatomic exchange calculation method for non-collinear spin systems, we have analysed the exchange interactions and magnetic moments of Co with fcc, bcc and hcp crystal structures. We found that elemental Co is unique in that it has excitations energies that reflect an almost perfect Heisenberg system \amendyk{in a rather wide range of angles between the spins}. This has a significant importance in the correct description of the time evolution of the atomic magnetic moments under the influence of a temperature-dependent laser pulse. Note that in contrast to Co, bcc Fe shows significant spin configuration-dependence, for any definition of the spin-Hamiltonian, as it was shown here and in previous studies\cite{attila}. 
 
\amendyk{Mn spins, on the contrary, exhibit strongly non-Heisenberg behavior already for small degree of inter-atomic non-collinearity. A relatively small amount of Mn dopants, as in the case of the alloys studied here, is not expected to drastically alter the system's properties. We note, however, that high Mn concentration will definitely lead to the breakdown of the Heisenberg picture and we plan to investigate it in detail in future.}

The calculated structural and magnetic properties of Co-rich Co-Mn alloys are compared with experimental data. We find that they are in very good agreement with observations. The calculated $T_{\rm C}$'s reproduce well the measured values and shows a linear decrease as a function of increasing Mn content, in line with the experiments. The magnon dispersion curves of fcc, bcc and hcp Co are plotted along the high symmetry directions of the Brillouin zone and they are indeed in good agreement with experimental data, where comparison can be made.  

We have also addressed the temporal behavior of the magnetism of Co in the bcc, fcc and hcp structures as well as Co-Mn alloys, after a laser excitation. Ultrafast magnetization dynamics of these Co systems were studied for different damping parameters, and it was found that the demagnetization behaviour depends critically on the damping parameter as well as the strength of the exchange interaction represented by different concentrations of Mn. This becomes especially interesting for Co-Mn alloys that have very different values of the damping parameters and exchange interactions of the constitute, which lead to drastically different magnetization dynamics of the Co and Mn sublattices. 

\section{Acknowledgements}
The authors thank to Tom Silva, Justin Shaw, Martin Shoen and A.N. Yaresko (MPI Stuttgart) for fruitful discussions. We acknowledge financial support from the Swedish Research Council. O.E Acknowledged support from KAW (projects 2013.0020 and 2012.0031). O.E. and E. K. D.-Cz. acknowledges STandUP for financial support. The calculations were performed at NSC (Link\"oping University, Sweden) under a SNAC project.

\end{document}